\documentclass{aa}
\include{epsf}
\def\lesssim{\mathrel{\hbox{\rlap{\hbox{\lower4pt\hbox{$\sim$}}}\hbox{$<$}}}}

\def\gtrsim{\mathrel{\hbox{\rlap{\hbox{\lower4pt\hbox{$\sim$}}}\hbox{$>$}}}}

\usepackage{graphics}

\renewcommand{\apj}{ApJ}
\renewcommand{\apjl}{ApJL}
\renewcommand{\apjs}{ApJS}

\renewcommand{\aap}{A\&A}

\newcommand{\etal} {et al.}
\newcommand{\be}{\begin{equation}}
\newcommand{\ee}{\end{equation}}
\newcommand{\ba}{\begin{eqnarray}}
\newcommand{\ea}{\end{eqnarray}}
\newcommand{\Ex}{\cdot 10^}
\begin{document}
\title{Axi-symmetric models of ultraviolet radiative transfer with
applications to circumstellar disk chemistry}
\author{Gerd-Jan van Zadelhoff \inst{1} \and Yuri Aikawa \inst{2} \and
M.R. Hogerheijde \inst{3} \and Ewine F. van Dishoeck \inst{1}}

\offprints{G.J. van Zadelhoff,\email{zadelhof@knmi.nl}} 
\institute{Leiden Observatory, P.O. Box 9513, 2300 RA
Leiden, The Netherlands  \and Dept. of Earth and Planetary Sciences,
Kobe University, 1-1 Rokkoudai, Nada-ku, Kobe, 657-8501, Japan \and Steward
Observatory, The University of Arizona, 933 N. Cherry Ave.  Tucson AZ,
USA}
\date{Received **** / Accepted ....}

\abstract{A new two-dimensional axi-symmetric ultraviolet radiative
transfer code is presented, which is used to calculate 
photodissociation and ionization rates for use in chemistry models of
flaring circumstellar disks. Scattering and absorption of photons
from the central star and from the interstellar radiation field are
taken into account. The molecules are effectively photodissociated in
the surface layer of the disk, but can exist in the intermediate,
moderately warm layers.  A comparison has been made with an
approximate 2D ray-tracing method and it was found that the latter
underestimates the ultraviolet field and thus the molecular
photodissociation rates below the disk surface. The full 2D results
show significantly higher abundances of radicals such as CN and C$_2$H
than previous work, partly due to the fact that CO is dissociated to
greater depths. Results for different stellar radiation fields are
also presented. The CN/HCN ratio shows a strong dependence on the
stellar spectrum, whereas other ratios such as HCO$^+$/CO show only
little variation.  { }} \titlerunning{Chemical effects due to 2D UV
transfer} \authorrunning{van Zadelhoff \etal}

\maketitle

\section{Introduction}
Numerous recent observations of dust and molecular line emission have
increased the understanding of disks around isolated T-Tauri stars in
both their appearances and internal structure (Beckwith \& Sargent
1996, Dutrey et al.\ 2000, Qi 2001). The modeling to date can be
divided in two categories.  A first category deals with the physical
structure of disks. It was proposed by \cite{hartman} that disks have
a flared geometry, where the stellar light puffs up the outer regions
of the disk. Recent papers by Chiang \& Goldreich (1997,\ 1999),
\cite{1997A&A...318..879M}, D'Alessio et al. (1998,\ 1999), 
Dullemond et al.\ (2002) and Nomura (2002) show similar results by
calculating explicitly the density and temperature distributions in
disks in hydrostatic equilibrium. Chiang \& Goldreich (1997) fit
spectral energy distributions by considering disks with a bi-layered
temperature structure. This structure is calculated assuming that the
ultraviolet and visible light from the star heats up the outer layer
of the disk, and does not effect the shielded region
directly. D'Alessio et al.\ (1998) calculate the full temperature
structure in the vertical direction and include the effects of
accretion and viscous heating.

The second category takes a physical disk model and uses a time-dependent
chemistry code to calculate the molecular abundances in the gas 
(Aikawa et al. 1997; Aikawa \& Herbst 1999; Willacy et al. 1998;
Willacy \& Langer 2000). 
Aikawa \& Herbst (1999) use the minimum-mass solar nebula model of
Hayashi (1981), whereas Willacy et al. (2000) use the Chiang \&
Goldreich model. In the first model, the temperatures are very low
throughout the disk, thereby freezing out the molecules onto grains,
while in the latter model nearly all molecules are dissociated due to
the ultraviolet light in the upper layer and are depleted in the cold
lower layer. Both models have to assume either an artificially low
sticking probability or a very high photo-desorption rate to keep
molecules in the gas.

In the latest paper by Aikawa et al.\ (2002, hereafter Paper I), the
D'Alessio et al. model was used to calculate molecular abundances.
The stellar and interstellar ultraviolet (UV) radiation heats the dust
and gas, and dissociates molecules in the surface region of A$_{\rm
v}\lesssim 1$ mag.  In the deeper layers, visible photons with lower
energies are available to heat the grains and gas, but there are much
fewer ultraviolet photons for dissociation than in the surface
layer. The dust in the surface layer, heated by the ultraviolet and
visible photons, in turn emits infrared photons, which can travel even
further toward the mid-plane until they are eventually absorbed.  
Viscous heating is only important in the inner few AU of the disk,
which are not considered in this study. As a result, the disk has
three distinctive chemical layers.  In the surface layer with
temperatures $T >40 $ K and density $n_{\rm H}\sim 10^{4}-10^{5}$
cm$^{-3}$, molecules are mostly dissociated.  In the intermediate
layer, conditions are comparable to those in dense interstellar
clouds, i.e., the temperature is 20--40 K and density is $\sim
10^{6}-10^7$ cm$^{-3}$. This layer contains sufficient amounts of
gaseous molecules to reproduce the observations of some disks.  In the
innermost layer including the midplane, the density is high ($n_{\rm
H} > 10^7$ cm$^{-3}$) and temperature is low ($T\lesssim 20$
K). Nearly all the gaseous species, except for H$_2$ and He, freeze
out onto the grains and stay there until the grains are again heated,
e.g. when most of the disk has dissipated and becomes optically thin
to ultraviolet radiation or when the grains are dynamically
transported to a warmer part of the disk.

In this work, we again adopt the density and temperature distribution
of the D'Alessio et al. model, and use a new two-dimensional (2D)
ultraviolet radiative transfer code to describe the radiative effects
on the chemistry.  Previous works, including our Paper I, treated the
UV and optical radiative transfer in disks in a one dimensional (1D)
or approximate 2D method.  In the former case it is assumed that the
disks are geometrically thin, i.e. the scale-height of the disk is
small compared to the distance to the star, and therefore radiative
transfer in the radial direction is negligible compared to that in the
vertical direction (\cite{1997ApJ...490..368C}, D'Alessio et
al. 1998).  In the latter case the attenuation both along the line of
sight from the star and in the vertical direction is calculated (Paper
I).  This method cannot treat scattering effects in principle and
therefore only gives an approximate depth-dependent intensity; forward
scattering will help stellar photons at UV and optical wavelengths to
penetrate deeper into the disk, which will affect the chemistry
through photodissociation. One important aim of this paper is a
direct assessment of the effects of the correct 2D treatment compared
with the earlier more approximate methods.  The 2D radiative transfer
code also enables us to consider dissociation of H$_2$ and CO via the
stellar UV radiation, which was neglected in Paper I. In addition the
dependence of molecular abundances and line intensities on the stellar
spectrum can be studied.

Paper I and this paper are complementary to the recent work of
Markwick et al.\ (2002). Whereas our studies focus on the outer
regions of the disk ($>$50 AU), Markwick et al.\ (2002) consider the
inner 10 AU in a  quasi-static treatment; the chemistry is
calculated locally, with the starting conditions defined by the
chemical abundances calculated at more distant radial points.  Their
calculations confirm the results of earlier 1D studies that the
chemistry in the inner region is driven by thermal desorption.  The
inclusion of X-rays in their model shows that radicals such as CN and
C$_{2}$H are important tracers of the ionization in disks, a
conclusion also reached for the outer disks in our papers.

The outline of this paper is as follows. In \S \ref{sec: chemsec}, the
coupling between the radiation and chemistry is discussed. In \S
\ref{sec: resultsec} the results are presented of the inclusion of
continuum radiative transfer on the chemistry and compared with the
approximate 2D method. The differences in line strength of the various
molecules are calculated in \S \ref{sec: linessec}.  Discussion
is in \S \ref{sec: discussec}, followed by a conclusion in \S 6.
Finally the adopted 2D
Monte Carlo UV radiative transfer code is explained in detail in
Appendix \ref{sec: appenA}.
 
\section{Chemistry and ultraviolet radiative transfer}
\label{sec: chemsec}

\subsection{General considerations}

The gas-phase molecular abundances are governed by two different
effects of the radiation. On the one hand, the stellar radiation
heats up the disk, assuring that freeze-out onto grains does not
occur.  This heating by direct radiation is dominated by the
lower energy photons emitted by the cool central star ($T_*=4000$ K).
In this paper, we assume that the gas and dust temperatures are
coupled and adopt the calculated dust temperatures from D'Alessio et
al.\ (1999).  On the other hand, the higher-energy UV radiation
dissociates and/or ionizes molecular and atomic species and most T
Tauri stars have a relatively large UV flux compared to main sequence
stars of the same temperature  (e.g. Costa et al.\ 2000).  The mean
intensity of the visible and ultraviolet light at a position in the
disk is controlled by absorption and scattering, making the solution
non-trivial. In this section, we briefly discuss the UV radiative
transfer in the disk. The numerical details of the Monte Carlo code
are explained in more detail in Appendix \ref{sec: appenA}.

Before combining the radiative transfer with the chemistry, the
different time scales in the problem are checked. The three time
scales of interest are: the radiative time scale, the chemical
time scale, and  the dynamical time scale.  Since the current line
observations of disks have poor spatial resolution, even for sources
mapped with millimeter interferometers (Dutrey et al.\ 1996, Qi 2001),
they are mostly sensitive to the outer regions of the disk. A
comparison of the time scales is therefore most appropriate at
distances $\ge$ 100 AU from the star.

The first time scale represents the time needed for the radiation
field to adapt to fluctuations in physical and/or chemical
processes. The disk itself is too cold to emit ultraviolet or visible
photons, therefore the mean intensity at each point in the disk
depends only on changes in the source, both stellar or interstellar,
and the density distribution in between the source and that point.
The radiation field reaches its mean value on the same time scale as
it takes the light to travel from the source to the point of interest;
for 100 AU this is $\sim 14$ hours from the star.  So the
radiation field is always in equilibrium in the time scales we are
concerned with in this paper.

The chemical time scale, i.e., the time in which the number density of
a species reaches its steady-state value, varies with molecular
species and position in the disk. In the intermediate molecular
layer we are concerned with, it is of the order of $10^{4}$--$10^{5}$
years for many organic molecules (Paper I).  On the other hand, in the
midplane layer the time scale of adsorption, which is the dominant
process, is very short: $\sim 10^2\, (10^8$ cm$^{-3}/n_{\rm H})$ yr.

 There are several dynamical time scales in the disk. The typical
accretion time scale of the disk as a whole is estimated to be $10^6$
yr, derived from dividing the disk mass ($10^{-2}$ M$_{\odot}$) by the
mass accretion rate ($10^{-8}$ M$_{\odot}$ yr$^{-1}$). The local
accretion timescale at $R=100$ AU, i.e. the time it takes for a parcel
of gas to move 5 AU (less than the radial gridsize in our model)
radially inwards, is estimated to be shorter, $7\times 10^4$ yr from
the local radial drift velocity $v_{\rm R}=\dot{M}/[2\pi R
\Sigma(R)]$.  Since accretion is caused by the transport of angular
momentum due to turbulence, the timescale for turbulent mixing is also
relevant. This timescale for radial mixing over $\triangle R$ is $\sim
(\triangle R)^2/(l_{\rm turb} v_{\rm turb})$, for $\triangle R$ larger
than the size of turbulent eddy $l_{\rm turb}$ (e.g. Aikawa et
al. 1996). The actual Eddy size ($l_{\rm turb}$) and turbulent
velocity ($v_{\rm turb}$) are highly uncertain, but if they are
assumed to be 10\% of the vertical scale height and sound velocity,
the timescale for global mixing (i.e. $\triangle R \sim R \sim 100$
AU) is of the order of $10^6$ yr. The local mixing timescale with
smaller $\triangle R$ (or $\triangle Z$) is accordingly shorter. 

In this paper we assume that the disk is static; the molecular
abundances are calculated at each point ($R, Z$) in the disk, without
considering any accretion flow or mixing for simplicity. Although this
static model is invalid from a physical point of view, as the
comparison of the different local time-scales show above, this
simplification is adopted for the following two reasons.  First, disk
chemistry is a complicated mixture of various chemical processes and
physical processes (e.g., density and temperature distributions,
radiative transfer and hydrodynamics).  In this series of papers, the
importance of each of the processes is investigated step by step; in
Paper I the chemical processes which are important in the different
regions of the disk were analyzed, and in the current paper the effect
of 2-dimensional UV radiative transfer is investigated. Mixing and
accretion smooth out the molecular stratification over some length,
determined by the balance between the chemical and dynamical time
scales. At the same time radical species are transfered in the
intermediate layer, activating chemical reactions in neighbouring
regions. These effects are expected to be larger in the vertical
direction than in the radial direction, since the radial distributions
do not vary much in the outer disk between 50 and 400 AU (Paper I).
They should be investigated in detail in independent future work.
Second, full coupling of these effects in 2- or 3-dimensional
calculations would be very complicated and too time-consuming with
current computational facilities. In order to accomplish the full
coupling of chemistry, radiative transfer and (magneto-)hydrodynamics,
the chemical reaction network needs to be significantly reduced by
carefully choosing the dominant chemical reactions in each region of
the disk. The results presented in our papers using a large chemical
network will assist the construction and testing of future coupled
chemical-hydrodynamical models.

Another relevant physical process is the settling of dust and
grain-growth.  In this paper we assume for simplicity that the optical
properties of dust grains are the same as those of interstellar
clouds, and that the gas and dust are well mixed.  Settling of small
(sub-micron sized) grains, which are the main source of the UV
opacity, takes place over a period of at least 10$^{6}$ years
(e.g. Nakagawa, Nakazawa, \& Hayashi 1981; Weidenschilling 1997). The
coagulation timescale, however, is much shorter. In fact, the spectral
energy distributions (SEDs) of T Tauri stars indicate significant
increase to mm-sized grains compared with interstellar dust (Chiang et
al.  2001; D'Alessio, Calvet, \& Hartmann 2001). Inclusion of dust
coagulation in our radiation-chemistry disk model is however more
complicated, since variations in dust-properties, scattering and
possible decoupling of the gas and dust temperatures could be of
essential importance. In previous work on the vertical disk structure,
the temperature distribution is determined by radiative transfer, with
dust as the main source of opacity, whereas the density distribution
is determined by the hydrostatic equilibrium, assuming the same gas
and dust temperatures. If the number of small grains decreases due to
coagulation, the two temperatures may decouple if not enough
collisions between gas and dust particles take place.  Calculations of
the decoupled gas and dust temperature have only been performed for
tenuous disks which have dust masses that are at least two orders of
magnitude less than the disks studied in this work and are optically
thin to UV and optical radiation, simplifying the radiative transfer
(Kamp \& van Zadelhoff 2001). This problem of dust coagulation is left
for future work, and a basic 2-D radiation-chemistry model with
interstellar-sized grains is established in this paper.

The chemical model adopted in this paper is the ``new standard model''
(NSM) for the gas-phase chemistry (Tervieza \& Herbst 1998; Osamura et
al.\ 1999), extended to include deuterium chemistry, adsorption of
molecules onto grains, and thermal desorption (Aikawa \& Herbst 1999;
and references therein). Chemical processes by X-rays are not included
in this paper.

\subsection{Impact of ultraviolet radiation}

The chemical abundances are related through a reaction network. In
this network the dissociation- and ionization-rates $k_{i}$ of the
different species are calculated using \begin{eqnarray}
k_i=\int^{\infty}_{0} 4 \pi \frac{\lambda}{hc} \sigma_{i}(\lambda)
J_{\lambda} d \lambda
\label{eq: rate1}
\end{eqnarray}
where $h$ is the Planck constant, $c$ the velocity of light and
$J_{\lambda}$ the mean intensity at wavelength $\lambda$, defined as
\begin{eqnarray} J_{\lambda}=\frac{1}{4 \pi} \int_{\Omega} I_{\lambda} \, d \Omega.
\end{eqnarray} 
Some molecules can be dissociated by absorption of photons over a
continuous part of the spectrum, while others need discrete line
transitions or a combination of the two (van Dishoeck 1988).  The
cross sections $\sigma_{i}(\lambda)$ or oscillator strengths
$f_{i}(\lambda)$ need to be specified for each transition for each
molecule. We adopt here the values given by van Dishoeck (1988),
updated by Jansen et al.\ (1995a \& 1995b). For species for which no
data on the cross sections are available, a rate given by a similar
type of molecule was adopted.
 
The photodissociation rate of a species depends on the mean intensity
of the radiation in the spatial grid, which is obtained via the 2D
Monte Carlo code. This code is based on the same principles as those
developed by \cite{boisse}, Spaans (1996) and Gordon et al.\ (2001)
for interstellar applications and e.g., Whitney \& Hartmann
(1992) for circumstellar disks. In general, one needs to define a
source with a surface area $S$, where the total energy is the flux
from the source times the surface area.  The total energy is divided
into a number of photon-packages ($N_{\lambda}$), which then travel
through the grid. The energy density at each point is calculated using
both the number of photons passing through a grid cell and the mean
energy transported by each photon during its transfer. Photon packages
are either absorbed or scattered by dust grains and are followed until
they leave the system or have reached a minimum intensity.  Optical
properties of dust grains are assumed to be the same as those of
interstellar dust; the scattering is described by the
Henyey-Greenstein probability function (Eq. \ref{eq: henyey}), and
albedo and opacities are taken from Roberge et al. (1991).

In the Monte Carlo method, the radiation field within the disk is
 readily obtained for irradiation by both stellar and interstellar
 radiation: since each source of radiation gives its own energy
 density, calculations for different sources can be added to give a
 total energy density. In addition, radiation fields for different
 stellar spectra are easily calculated (Sec.\ 3.4).  The energy
 densities in the disk are directly proportional to the emitted
 stellar flux, and are thus scalable. Since the disk structure is
 independent on the precise UV spectrum of the central star,
 a single calculation is sufficient for obtaining the full UV
 radiation distribution for each assumed spectrum.

In the 912-1110 \AA\, region, additional attenuation due to
$\rm{H_{2}}$ and CO self-shielding was included. The shielding of
other species by $\rm{H_{2}}$ was ignored in these calculations. This
could lead to an overestimate of dissociation rates for species which
dissociate only in the 912-1110 \AA \, region by at least 20 \%
(Draine \& Bertoldi 1996). Self-shielding of H$_2$ and CO depends on
the column densities of the molecules themselves along  the photon
trajectories as well as the dust attenuation ($A_{\rm v}$), which means that
the chemical abundances and the UV radiative transfer need to be
solved simultaneously. For the interstellar UV field this is easily
done via a 1D slab model (van Dishoeck \& Black 1988, Lee et
al. 1996), which is adopted in this paper and previous works
(e.g. Paper I). Shielding of the stellar UV radiation is more
difficult to calculate, since iteration of the chemistry and 2D
radiative transfer is very time-consuming.  Since we are not
interested in the precise position of the $\rm{H-H_{2}}$ transition,
this iteration is performed only once: we first calculate the mean
intensity assuming that all hydrogen is in molecular form and adopt
the shielding factor given by Draine \& Bertoldi (1996), their formula
37 (\ref{eq: bert1}, see Sec. \ref{sec: shield} for details), and then
obtain the H$_2$ abundance from chemical network.  The self-shielding
of CO has been calculated in a similar way using the shielding
functions from Lee et al.\ (1996).

\section{Results}
\label{sec: resultsec}
\subsection{UV radiation and photodissociation rates}

The temperature and density structure of the disk model by D'Alessio
et al.\ (1999) is adopted as our fiducial model, which was calculated
for an accretion rate $\dot{M}=10^{-8} M_{\odot}$ yr$^{-1}$ and
viscosity parameter $\alpha=0.01$. The disk has a radius of 400 AU and
a mass of 0.05 M$_{\odot}$, representative of the disk around the T
Tauri star LkCa15 for which molecular data exist (Qi\ 2001, van
Zadelhoff et al. 2001, Thi et al.\ 2002). Similar sizes or even
larger have been observed for a number of gas disks, see, e.g., Simon
et al.\ (2001): LkCa 15 (650 AU; 435 AU by Qi 2001), DM Tau (800 AU),
DL Tau (520 AU) and GM Aur (525 AU).  These are the same disks for
which observations of molecules other than CO are available and for
which chemical models have been presented in Paper I.  The adopted
disk model is gravitationally stable according to the Toomre criterion
up to at least 340 AU (D'Alessio et al.\ 1999, see also Bell et al.\
1997 for a discussion on this point). We therefore truncated our disk
at the radial grid cell in which this radius is reached, $R_{\rm
out}$=400 AU.

  The disk structure is calculated self-consistently for a
stellar temperature of 4000 K and $R_*$=2 $R_{\odot}$ (D'Alessio et
al. 1999).  The disk is assumed to be subjected to UV radiation from
the interstellar radiation field and from the star. The Draine (1978)
field is used to represent the interstellar field between 912 and 2000
\AA, extended to longer wavelengths as specified by van Dishoeck \&
Black (1982).  For the star, initially the same spectral shape is
taken, but scaled by a factor such that the intensity at $R=100$~AU
incident on the surface of the disk is a factor of $10^4$ higher than
the interstellar field. This choice was motivated by Herbig \&
Goodrich (1986) and was also adopted in Paper I. In Sect.\ 3.4,
different assumptions for the stellar radiation field are explored.
It is assumed that the additional UV radiation does not change the
disk structure. The integrated flux in the range 912--4000 \AA,
-- the range important for the chemistry--, can contribute to a
maximum of 10\% of the total energy for the stars modeled in this
paper.  The UV radiative transfer was calculated using the grain
properties and extinction curve by Weingartner \& Draine (2001).

\begin{figure*}[ht!]
\resizebox{\hsize}{!}{\includegraphics{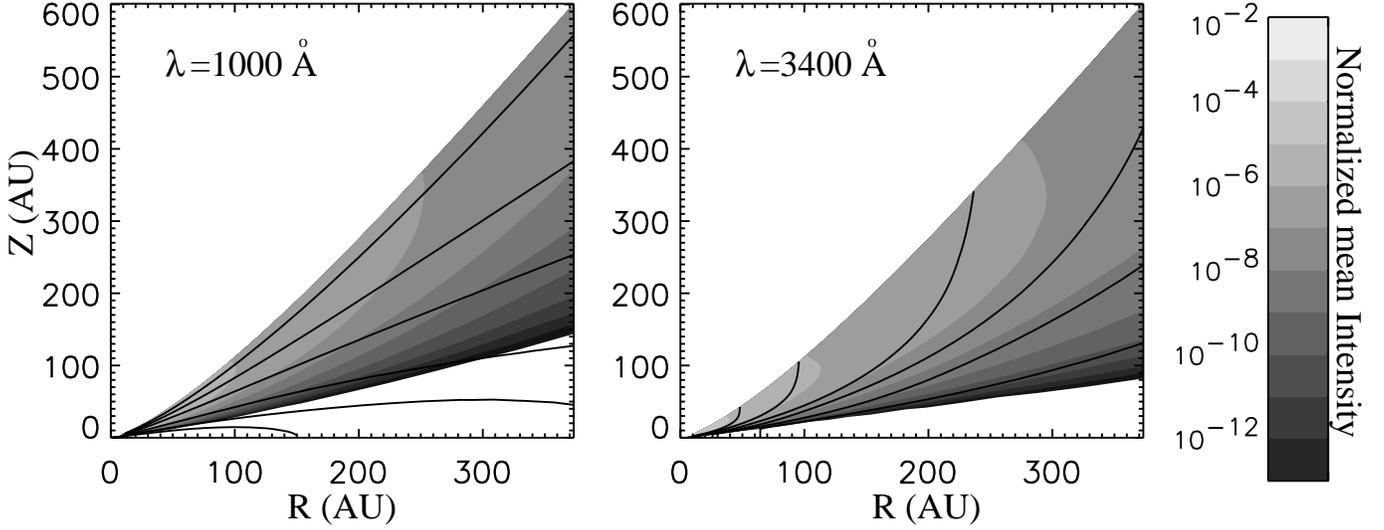}}
\caption{Normalized mean intensities at 1000 and 3400 \AA \,for the
adopted model. The mean intensities have been normalized to the value
in the innermost cell to better visualize the effects of grain
scattering and absorption. Overplotted are the H$_{2}$ number density
contours (from surface to midplane: 5$\times10^{4}$, 2$\times10^{5}$,
10$^{6}$, 10$^{7}$, 10$^{8}$, 10$^{9}$) [cm$^{-3}$] on the left and
the dust temperature contours (upper left to lower right: 90, 70, 50,
40, 30, 20) [K] on the right. Only one quadrant of the 2D flaring disk
is shown.}
\label{fig: radpap6}
\end{figure*}

\begin{figure}[ht!]
\resizebox{\hsize}{!}{\includegraphics{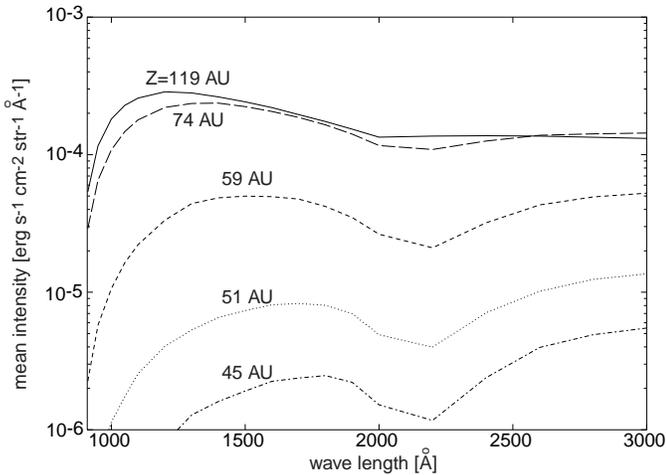}}
\caption{Spectrum of the mean intensity $J_{\lambda}$ at different
heights in the disk for a radius of $R=105$ AU.  Numerals in the
figure indicate the height from the midplane. The stellar UV is
assumed to be 10$^{4}$ times higher than interstellar field at $R=100$
AU and have the same shape.}
\label{fig: spect}
\end{figure}

Figure \ref{fig: radpap6} shows the resulting distribution of the UV
intensity in the disk for two different wavelengths, with overplotted
the density and temperature structure.  It is seen that for longer
wavelengths, the radiation penetrates deeper into the disk because of
the reduced absorption.  Figure \ref{fig: spect} shows the UV spectrum
at various heights at $R=105$ AU.  Even at short wavelengths (912
--2000 \AA), the intensity is still comparable to the unattenuated
interstellar radiation field down to 35 AU at $R\approx100$ AU.  The
longer wavelengths visible photons start to dominate the radiation
field below 50 AU.  The dip at $\lambda \sim 2200$ \AA \, is caused by
the bump in the extinction curve.

\begin{figure}[ht!]
\resizebox{\hsize}{!}{\includegraphics{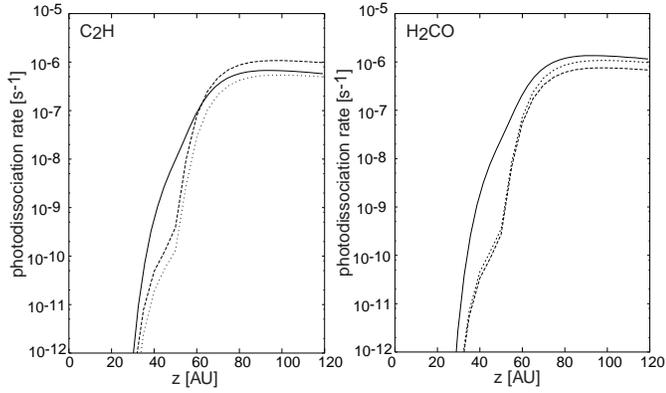}}
\caption{Photodissociation rates of C$_2$H and H$_2$CO as functions of
height from the midplane at $R=105$ AU. Rates are calculated using the
radiation field obtained via the full 2D Monte Carlo radiative transfer
({\em solid lines}). For comparison, rates obtained via the
approximate 2D method adopting the depth-dependent function of van
Dishoeck (1988) ({\em dotted lines}) and the new standard model ({\em
dashed lines}) are shown.}
\label{fig: disso}
\end{figure}

The effect of radiation on the molecular abundances is reflected in
the different dissociation or ionization rates.  For example,
Fig. \ref{fig: disso} shows the dissociation rates of C$_2$H (C$_2$H
$\to$ C$_2$ + H) and H$_2$CO (H$_2$CO $\to$ CO + H + H).  C$_{2}$H is
dissociated by lines in between 1080--1530 \AA \, and H$_{2}$CO is
dissociated due to both line (1100 -- 1749 \AA) and continuum radiation
(912--1505 \AA). The solid lines show the rates calculated from the
stellar and interstellar radiation field, obtained via the full 2D
radiative transfer. The dotted and dashed lines show the rates
calculated using the depth-dependent dissociation rates of van
Dishoeck (1988) and NSM, respectively. Van
Dishoeck (1988) calculated a plane-parallel PDR model, and obtained
the dissociation rates $k$ as functions of depth from the cloud
surface. These were fitted by the single exponential decay function:
\begin{equation}
{\it k}={\it k_{0}}\rm{e}^{-\beta A_{\rm v}}.
\label{eq: decay}
\end{equation}
 The NSM assumes the same function, but with slightly different
coefficients. In Aikawa \& Herbst (1999) and
Paper I, the visual extinction $A_{\rm v}$ from the central star is
obtained at each position in the disk by integrating the density
distribution along the line of sight to the central star, and the
dissociation rate is obtained with the NSM depth-dependent rates.
The dissociation rate via the interstellar UV field is obtained in a
similar way using the depth-dependent NSM rates and calculating $A_{\rm v}$
from the disk surface. Figure \ref{fig: disso} shows the total
dissociation rate via UV radiation from the interstellar field and the
central star.  In the dotted and dashed lines, the two components can
be easily distinguished: radiation from the central star dominates at
$Z\gtrsim 50$ AU, while interstellar radiation dominates at $Z\lesssim
50$ AU.

In Fig. \ref{fig: disso}, it is seen that the dissociation rates are
significantly underestimated at $Z\sim 50$ AU, if the approximate
depth-dependent rates described above are used. This is caused by
geometrical effects: the depth-dependent rates are originally obtained
for a plane-parallel cloud, whereas the central star irradiates the
disk obliquely. For 50 \% of the photons which are scattered in the
disk surface, the grazing angle between the disk and the ray path
becomes larger than the initial value, which causes deeper penetration
into the disk.


\subsection{Vertical distribution of molecules}

\begin{figure}[ht!]
\resizebox{\hsize}{!}{\includegraphics{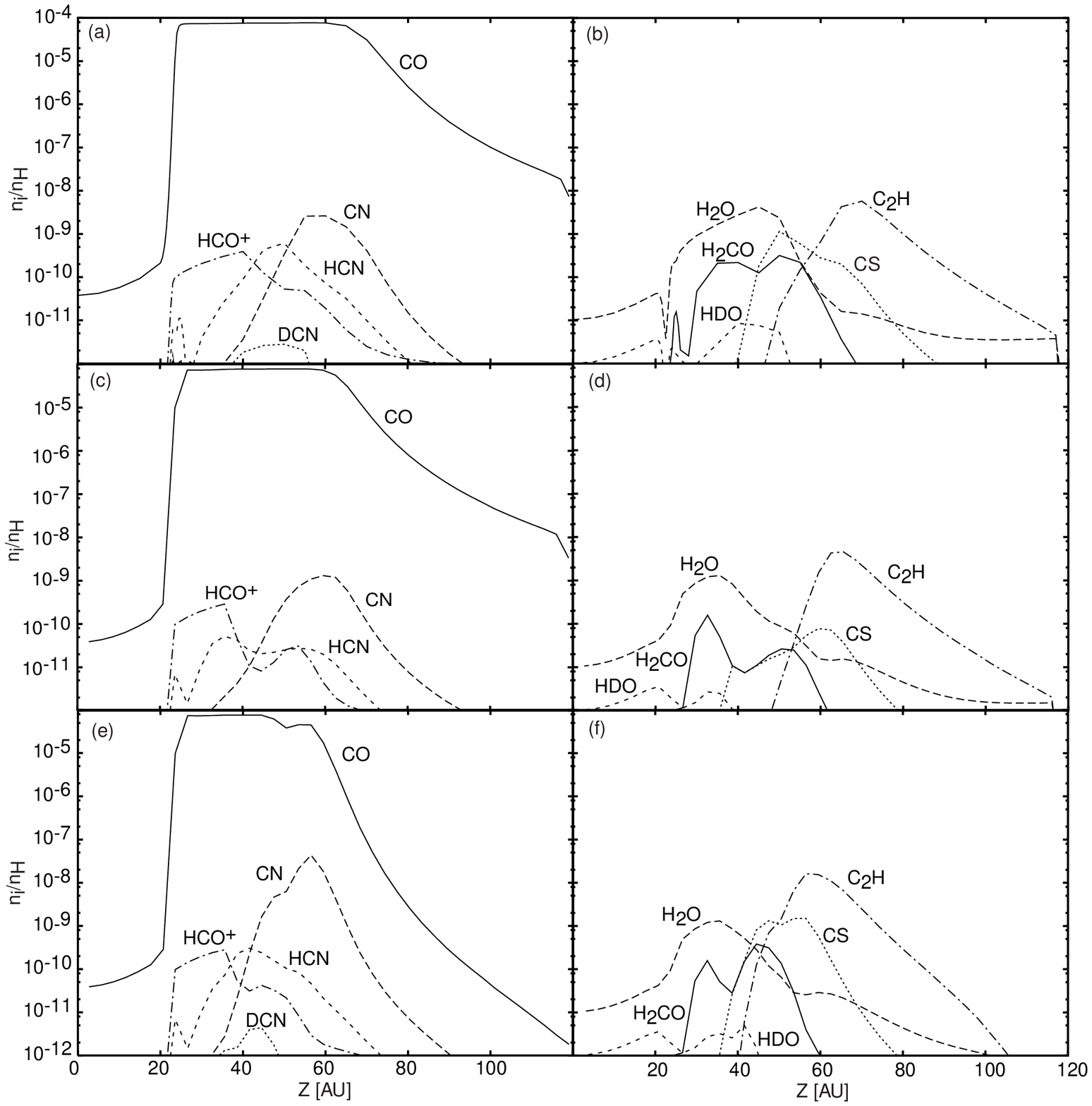}}
\caption{Vertical distribution of molecules at $R=105$ AU and $t=1\;
10^6$ yr.  In panels (a) and (b), the approximate-2D (A-2D) method is
adopted with the depth-dependent dissociation rates. The dissociation
rates used in panels (c) and (d) are determined using the full 2D
radiative transfer (ND-2D). In both cases the stellar UV is assumed
not to be energetic enough to dissociate H$_2$ and CO. In panels (e)
and (f), dissociation rates are determined using the full 2D radiative
transfer, and dissociation of H$_2$ and CO via the stellar UV is
considered.}
\label{fig: dist}
\end{figure}

Using the photodissociation rates calculated in Sec.\ 3.1, the
abundances of the molecules are calculated as a function of time.
Figure \ref{fig: dist} shows the vertical distribution of the
molecules at a disk radius of $R=105$ AU at $1\; 10^6$ yr. The
dissociation rates in panels (a) and (b) are obtained via the
depth-dependent function (Eq. \ref{eq: decay}), where the same model
is used as that presented in Paper I, but plotted at a different
radius. This model is called the A-2D (approximate 2D) model
hereafter. Two improvements are made in this paper: photodissociation
rates are calculated via the full 2D radiative transfer, and H$_2$ and
CO dissociation via stellar UV is included. To disentangle the two
effects, two models are run. In model ND-2D (no-dissociation,
panels c and d), the dissociation rates are obtained via the full-2D
radiative transfer results, but CO and H$_{2}$ are not dissociated by
the stellar light.  Comparison with A-2D (panels a and b) shows that
the CO abundances change only slightly, caused by the scattering and
further penetration of interstellar UV photons in full 2D.  Most other
species have lower abundances because the dissociation rate is
enhanced. The results obtained using the full 2D radiative transfer
and H$_2$ and CO dissociation via the stellar UV are presented in
panels (e) and (f). CO is dissociated even at $Z\sim 50$ AU in the 2D
model, while it is largely shielded from UV at $Z\lesssim 60$ AU in
the other two cases.
The peak abundances of radicals such as CN and C$_2$H (at $Z\sim 60$
AU), are enhanced, because the stellar UV dissociates CO to provide
more atomic carbon in the gas-phase reaction network. This zone
corresponds to the so-called ``radical zone'' found in the chemical
networks of PDRs (e.g., Jansen et al.\ 1995a \& 1995b, Sternberg \& Dalgarno
1995). The other species benefit from the CO dissociation as well;
their abundances are higher than in the ND-2D model.

\subsection{Column densities}
\label{sec: ondedtwod}

\begin{figure}[ht!]
\resizebox{\hsize}{!}{\includegraphics{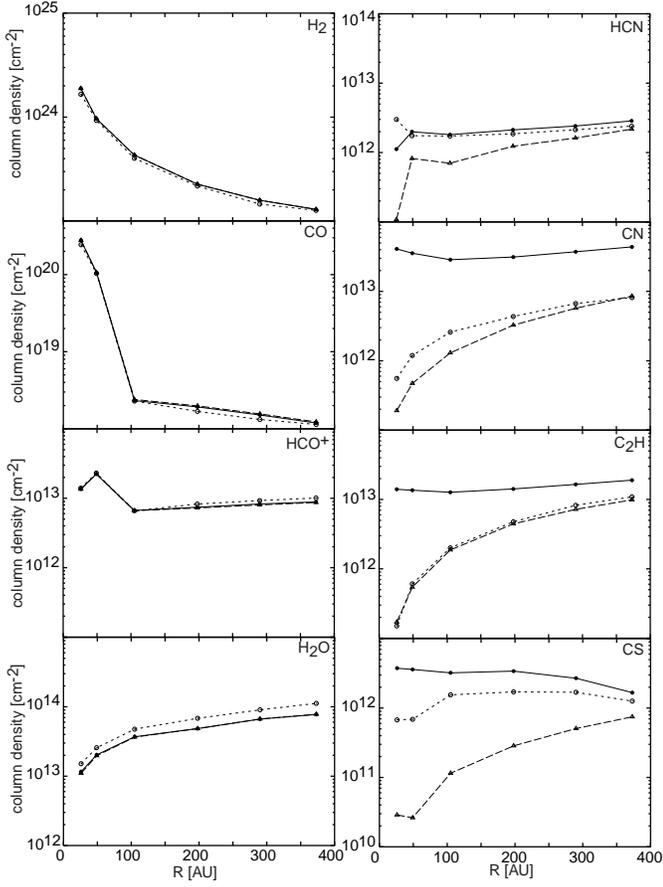}}
\caption{Radial distribution of column densities at $t=1\; 10^6$
yr. The dashed lines with open circles refer to the  A-2D model,
the long-dashed lines with the solid triangles to the ND-2D model,
and the solid lines with closed circles are for the full 2D model.}
\label{fig: column}
\end{figure}

The column densities obtained at each radius by integrating the
vertical distribution of molecules (Fig. \ref{fig: dist}) are
presented in Figure \ref{fig: column}. The column densities of
radicals (CN and C$_2$H) depend sensitively on the photodissociation
treatment, especially at the inner radii.  The column densities of CN
and C$_2$H in the A-2D and ND-2D models increase with radius because
of the lower density and lower flux of the destructive stellar UV in
the outer regions.  Inclusion of CO dissociation provides more carbon
in the gas-phase reaction network, which significantly enhances the
abundances of CN and C$_2$H.  At the inner radii, the higher
dissociation rates of these radicals are compensated by a larger
carbon supply. It suggests that an accurate calculation of the
radiative transfer of the stellar UV is important when considering the
abundances of radical molecules in disks.

Photodissociation of CO via the stellar UV at the disk surface,
however, does not affect the column density of CO significantly. It
should be noted that Fig. \ref{fig: dist} shows the molecular
abundances relative to hydrogen nuclei, and that the absolute number
density of CO is highest in the layer with $Z\sim 30$, which
contributes predominantly to the CO column density. Overall, column
densities of other species do not significantly depend on the
treatment of stellar UV. Close to the star, where the effect of the
stellar UV field is largest, differences can be seen. The HCN
abundance close to the star shows that 2D UV transfer is important.

As argued in Paper I and as can be seen in Fig. \ref{fig: dist},
molecules are abundant only in layers with certain physical
conditions.  Since the mass contained in those layers does not vary
much with radius, column densities of molecules such as HCN and
H$_2$CO show little dependence on radius. The column densities of CO
and HCO$^+$ abruptly change at $R\sim 100$ AU, inside of which the
temperature in the midplane is higher than the sublimation temperature
of CO ($\sim$ 20 K).

Molecular column densities were also calculated for disks with higher and
lower accretion rates of $10^{-7}$ and $10^{-9}$
M$_{\odot}$ yr$^{-1}$ and correspondingly larger and smaller disk
masses. For the same reasons stated above, the column
densities are similar to those in Fig. \ref{fig: column}, except for
CO and HCO$^+$ in the inner radius ($R\lesssim 100$ AU) and H$_2$, whose
column densities are higher (lower) in more (less) massive
disks (Paper I).

\subsection{Dependence on the stellar spectrum}
\label{sec: depenUV}

\begin{figure}[ht!]
\resizebox{\hsize}{!}{\includegraphics{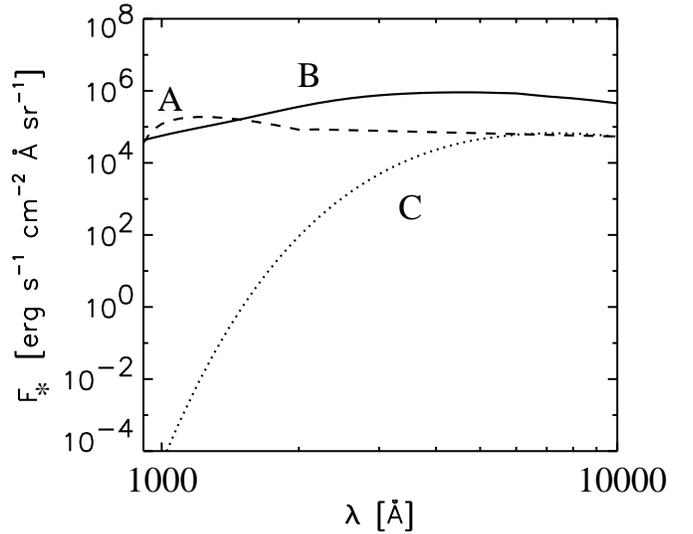}}
\caption{ Comparison of the mean intensity at a radius of $R=105$ AU
for the three stellar spectra: Spectrum A (dashed line), Spectrum B
(solid line), and Spectrum C (dotted line) (see text).}
\label{fig: radfieldcompare}
\end{figure}

So far, the UV spectrum from the central star has been assumed to be
the same as that of the interstellar radiation field, but with an
intensity $10^4$ times higher than the interstellar field at $R=100$
AU (hereafter referred to as Spectrum A). This spectrum was motivated
by Herbig \& Goodrich (1986) comparing IUE data in the wavelength
range between 1450 and 1850 \AA \ towards different T-Tauri stars. In
this subsection, the sensitivity of our results to the UV spectrum of
the central star is investigated by calculating two other models with
different stellar spectra: the observed spectrum of the T Tauri star
TW Hya (Spectrum B), and a black body spectrum at $T_{\ast}=4000$ K
without any excess UV (Spectrum C).  Costa et al.\ (2000) fitted the
UV continuum from TW Hya, observed with the IUE satellite, with a sum
of free-free plus free-bound emission at $3\times10^4$ K, plus a 4000
K black body spectrum for the star and a black body emission at 7900 K
covering 5 \% of the stellar surface.  We adopt his numerical data at
the stellar surface as Spectrum B. Note that no observational
constraints are available below 1200 \AA, especially in the critical
912--1100 \AA \, range where H$_2$ and CO are photodissociated.  It
will be important to obtain such data in the future with the FUSE
satellite. The third case with Spectrum C gives a lower limit to the
UV radiation from the central star, and may correspond to the
situation for disks around weak-line T Tauri stars (WTTS), although
the physical structure of such disks would be different from those
around classical T Tauri stars.  As in the full 2D model of the
previous section, the photodissociation rates are obtained via full 2D
radiation transfer, and dissociation of H$_2$ and CO via stellar UV is
taken into account. Photodissociation by the interstellar radiation is
included in all models. Figure \ref{fig: radfieldcompare} shows the
spectrum at the disk surface at $R\approx 100$ AU for the three
stellar models.

\begin{figure}[ht!]
\resizebox{\hsize}{!}{\includegraphics{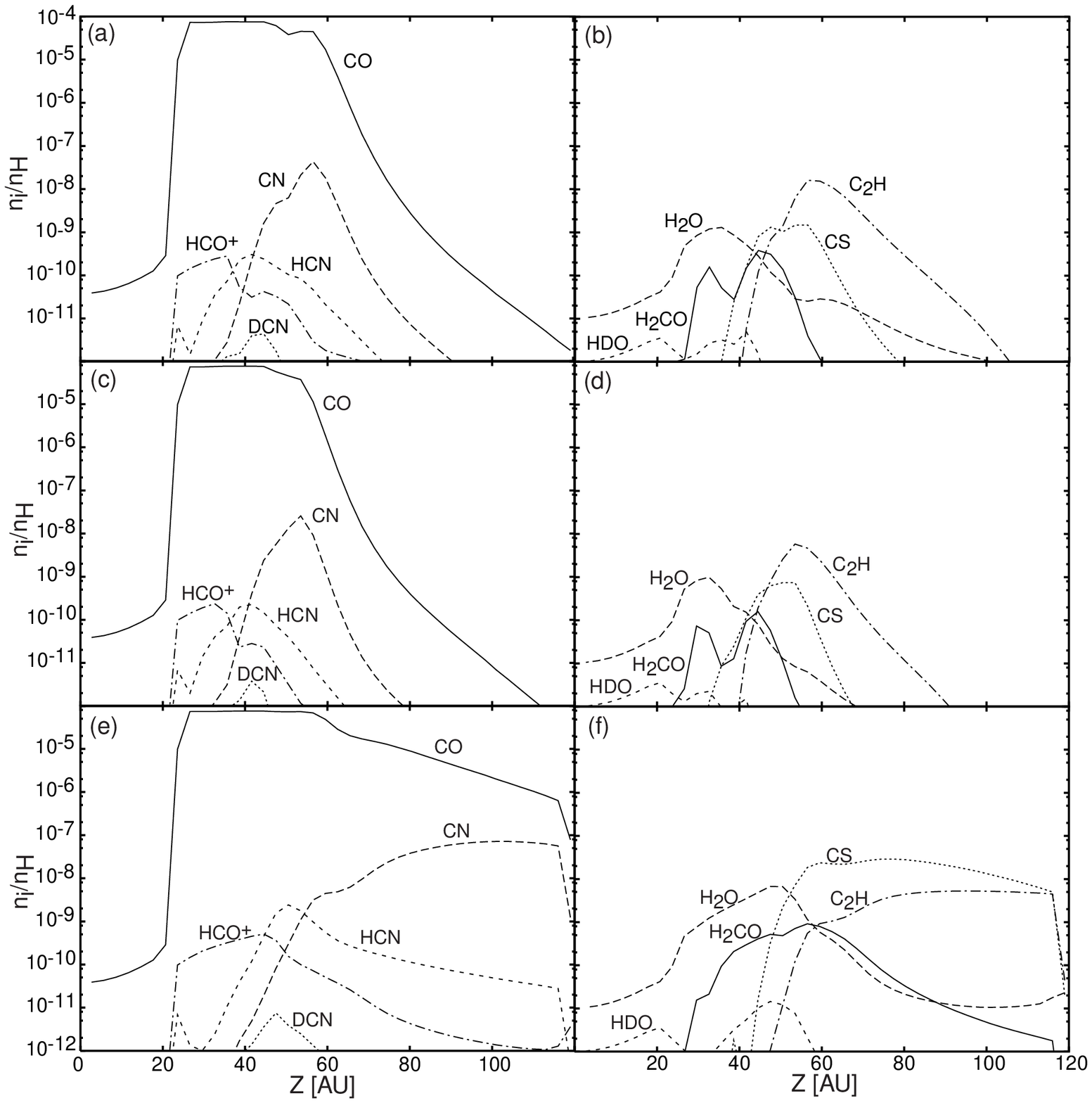}}
\caption{ Vertical distribution of molecules at $R=105$ AU and $t=1\;10^6$ yr.
Panels (a) and (b) are for Spectrum A, panels (c) and (d) are for Spectrum
B and panels (e) and (f) for Spectrum C}
\label{fig: dist_Costa4000}
\end{figure}

\begin{figure}[ht!]
\resizebox{\hsize}{!}{\includegraphics{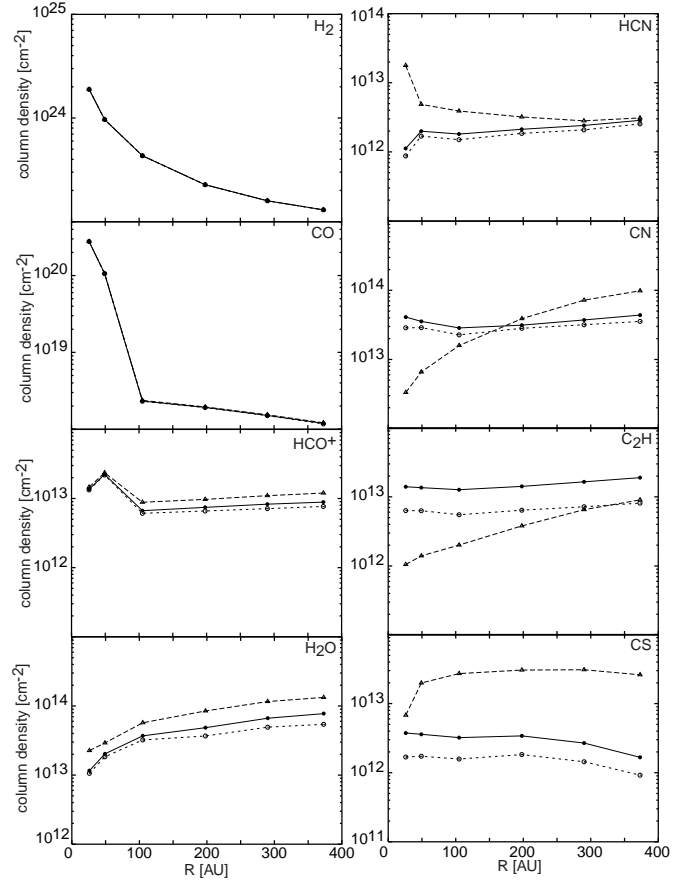}}
\caption{Radial distribution of molecular column densities at $t=1\;
10^6$ yr. Solid lines with closed circles are for Spectrum A, dashed
lines with open circles for Spectrum B, and long-dashed lines with
triangles for Spectrum C.}
\label{fig: column_Costa4000}
\end{figure}

Figures \ref{fig: dist_Costa4000} and \ref{fig: column_Costa4000} show
the vertical distribution of molecules at $R=105$ AU, and the radial
distribution of column densities for the three models with different
stellar UV spectra, respectively.  The model using Spectrum B shows
almost the same distributions and column densities as those obtained
with Spectrum A. This is not surprising since the spectrum of the
interstellar radiation field has a color temperature of $3\times
10^4$~K, close to that of the free-free and free-bound component
in Spectrum B. In the model without excess UV (Spectrum C), on the
other hand, the molecular distributions are far more extended toward
the surface layers.  The lower photodissociation rates at the disk
surface tend to enhance the column densities, but the radical (CN and
C$_2$H) column densities in the inner regions are lowered because of
the smaller amount of atomic carbon supplied via CO dissociation in
the intermediate layers.

\section{Molecular line emission}
\label{sec: linessec}
\subsection{Comparison of models} 

The comparison of our new results obtained via the full-2D radiation
transfer with the A-2D model and the ND-2D model shows some 
differences in molecular distributions, especially for radicals.

The question is if these differences are significant with the present
day observations or whether any differences are minimized after
solving the radiative transfer in the millimeter lines and convolving
the data with the telescope beam.  The CO and HCO$^{+}$
(sub-)millimeter lines are extremely optically thick in these
models. To get more information on the emission from deeper layers in
the disk, the radiative transfer has been run for $^{13}$CO and
H$^{13}$CO$^{+}$ as well. These species are assumed to follow the
abundances of CO and HCO$^{+}$ respectively, but their abundances are
lowered by a factor of 60, similar to the [$^{12}$C]/[$^{13}$C]
isotopic ratio in the solar neighborhood. This overestimates the
$^{13}$CO abundance for the full-2D case since CO is more
self-shielding compared to $^{13}$CO.

\begin{figure}[ht]
\resizebox{\hsize}{!}{\includegraphics{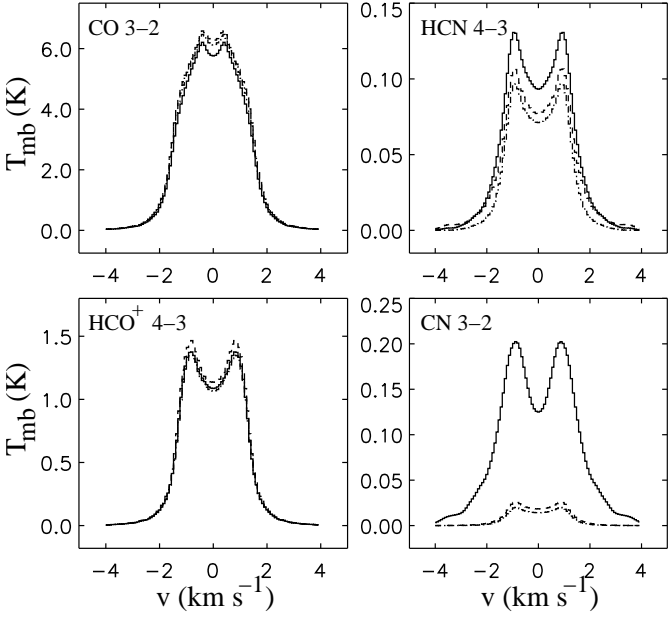}}
\caption{Comparison of emission lines for the full-2D radiative
transfer model (solid lines), the 2D radiative transfer model without
CO and H$_{2}$ dissociation (ND-2D; dot-dashed) and the approximate 2D
model (A-2D; dashed lines). The CO 3-2, HCO$^{+}$ 4--3, CN 3--2 and
HCN 4--3 lines are convolved with a beam equal to the size of the
source on the sky (5.33 \arcsec).}
\label{fig: onedtwod}
\end{figure}

As in Paper I, the level populations and emission lines of different
species are calculated using a 2D line radiative transfer Monte Carlo
code (Hogerheijde \& van der Tak 2000).  The calculations assume that
the 400 AU radius disk is seen at an inclination of 60$\degr$ at a
distance of 150 pc, appropriate for the case of LkCa15.  The results
are presented in two ways; first, the emission profiles for a few
characteristic molecules are plotted, and second, the ratios of
integrated intensities of the lines are compared in several tables.
In Fig.  \ref{fig: onedtwod}, lines are shown for CO 3-2, HCO$^{+}$
4--3, CN 3--2 and HCN 4--3. The CN 3--2 line is modeled assuming that
the three fine-structure lines are well shifted from each other, with
each fine-structure line a combination of three unresolved hyperfine
components. Therefore, the CN 3--2 line shows a regular double-peaked
profile instead of a combination of three double-peaked structures
summed at slightly shifted velocities.

\begin{center}
\begin{table}
\caption{\label{tab: tab1d2d} The ratios of the integrated intensities
for the full-2D, ND-2D and A-2D models (see text). The line emissions
have been convolved with a beam equal to the size of the disk on the
sky (5.33\arcsec).}
\begin{tabular}{llrcc}
\hline
Molecule & line & $\nu$ [GHz] &  full-2D & ND-2D \\
&  & &  $\overline{\rm A-2D}$ &  $\overline{\rm A-2D}$ \\
\hline
CO       & 6-5       &691.473 &0.88& 0.97\\
CO       & 3-2       &345.796 &0.92& 0.98\\
CO       & 2-1       &230.538 &0.93& 0.98\\
$^{13}$CO & 3-2      &330.587 &0.98& 1.03\\
$^{13}$CO & 1-0      &220.399 &1.05& 1.08\\
HCO$^{+}$& 4-3       &356.734 &0.94& 0.93\\
HCO$^{+}$& 1-0       & 89.189 &0.91& 0.89\\
H$^{13}$CO$^{+}$& 4-3&346.999 &0.94& 0.93\\
H$^{13}$CO$^{+}$& 1-0& 86.754 &0.98& 0.96\\
CN    & 3-2 &340.248 & 10.1 & 0.75\\
CN    & 2-1 &226.875 &5.81& 0.85 \\
HCN   & 4-3          &354.506 &1.21&0.81\\
HCN   & 1-0& 88.632 &1.19&0.85\\
\hline
\end{tabular}
\end{table}
\end{center}

 In Table \ref{tab: tab1d2d}, the ratios of the integrated intensities
of the full-2D and A-2D as well as the ND-2D over A-2D models are
given for a few lines. All models refer to a beam size
equal to the apparent size of the disk at a distance of 150 pc. In both
Fig.~\ref{fig: onedtwod} and Table \ref{tab: tab1d2d} the CO emission
is slightly lowered in the full-2D case due to the dissociation by
stellar UV.

Due to the dissociation of CO leading to atomic carbon, the CN
abundance and emission is enhanced up to an order of magnitude. This
is clearly seen in Fig.~\ref{fig: onedtwod}, where the A-2D and ND-2D
models have similar emission profiles, but full-2D is much higher. HCN
is linked to CN and its abundance is enhanced as well; however, in
contrast with CN, which can only be dissociated at wavelengths less
than 1000 \AA, HCN can be easily destroyed again by photodissociation,
resulting in both the abundances and line emission to rise only
mildly.  Among the three models, the ND-2D model gives the weakest HCN
emission.

\subsection{Dependence on stellar radiation field}
 
\begin{figure*}[ht!]
\resizebox{\hsize}{!}{\includegraphics{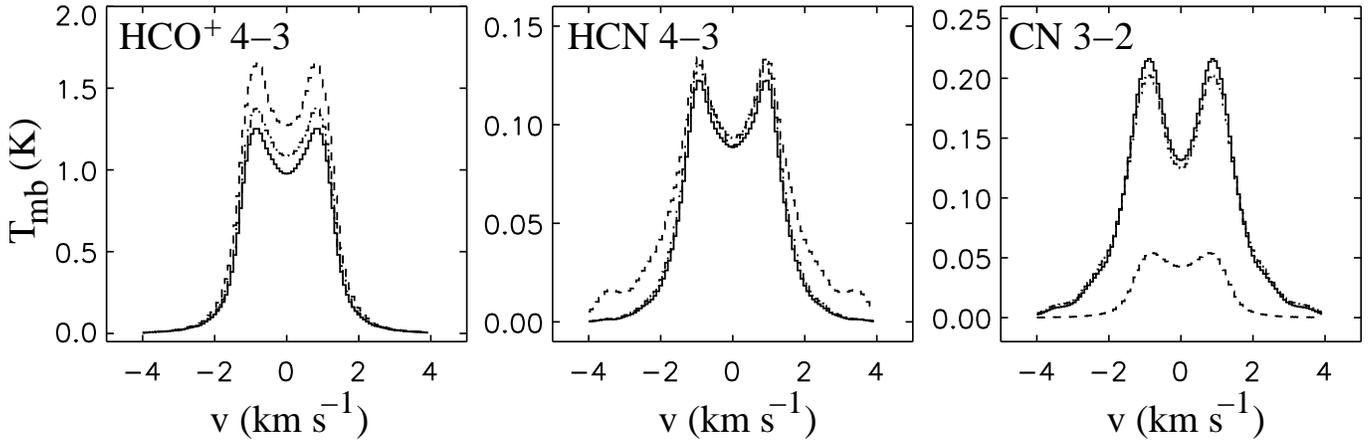}}
\caption{Comparison of HCO$^+$ 4--3 (left), HCN 4--3 (middle) and CN
3--2 (right) emission lines for the three stellar spectra.  The
dash-dotted lines are for Spectrum A, the solid lines for Spectrum B,
and the dashed lines for Spectrum C.  The spectra have been convolved
with a beam of 5.33\arcsec}
\label{fig: linesUVdepch6}
\end{figure*}

In the previous section it was shown that the chemical abundances need
a good description of the mean intensity of the UV radiation field.
The column densities of radicals such as CN are especially sensitive
to the UV spectrum. The models with excess UV radiation (Spectrum A
and B) have a flat column density distribution throughout the disk,
while the model without excess UV (Spectrum C) has an increasing CN
column density toward the outer radius.  Figure \ref{fig: linesUVdepch6}
shows line profiles of HCN, HCO$^{+}$ and CN for the three different
stellar spectra. The lines of HCO$^{+}$ nicely reflect the column
densities shown in Fig. \ref{fig: column_Costa4000}; the model with
Spectrum C gives the strongest emission.  The HCN column densities
have similar values at larger radii, which is reflected in the 4--3
emission lines. At smaller radii the column of the model with Spectrum C
dominates by a factor of 5. This is reflected in the higher emission in
the wings of the line profile. The excess emission at $|$v$| \sim$4 km
s$^{-1}$ is due to the undersampled grid in the inner region, but this
does not effect the integrated emission. The CN emission is discussed
in more detail in Sec. \ref{sec: nltechap6}.

\begin{table}
\caption{\label{tab: linesUVdep} Integrated intensities for the three stellar spectra (A,
B and C) convolved with a beam of 5.33\arcsec .}
\begin{tabular}{llccc}
\hline \multicolumn{2}{l}{molecular line} & A & B & C \\ 
 & & [K km s$^{-1}$]& [K km s$^{-1}$]& [K km s$^{-1}$] \\
\hline 
CO    & 6-5               & 11.19 & 11.79 & 12.65\\
CO    & 3-2               & 16.92 & 17.62 & 18.74\\
CO    & 2-1               & 17.86 & 18.52 & 19.58\\
$^{13}$CO    & 3-2        &  7.93 &  8.05 &  8.28\\
$^{13}$CO    & 2-1        &  7.84 &  7.93 &  8.09\\
$^{13}$CO    & 1-0        &  5.28 &  5.31 &  5.38\\
HCO$^{+}$    & 4-3        &  3.38 &  3.72 &  4.45\\
HCO$^{+}$    & 1-0        &  2.07 &  2.31 &  2.93\\
H$^{13}$CO$^{+}$ & 4-3    &  0.10 &  0.11 &  0.14\\
H$^{13}$CO$^{+}$& 1-0     &  0.04 &  0.05 &  0.06\\
CN    & 3-2               &  0.68 &  0.66 &  0.15\\
CN    & 2-1               &  1.94 &  2.00 &  1.01\\
HCN    & 4-3              &  0.35 &  0.37 &  0.46\\
HCN    & 1-0              &  0.56 &  0.63 &  0.77\\
\hline

\end{tabular}
\end{table}
 For a broader overview of the calculated emission lines, the
integrated intensities are given in Table \ref{tab: linesUVdep} for
several species and lines. These results were again calculated for the
source seen with a beam of 5.33\arcsec. The CO emission lines show
only a small dependence on the stellar UV field ($\approx$ 10\%),
where the case without excess UV (Specrum C) shows the highest
integrated flux.  The $^{13}$CO lines show an even smaller difference
since most of the emission comes from the warm intermediate layers and
is not affected by dissociation in the surface layer. HCO$^{+}$ and
its main isotopomer H$^{13}$CO$^{+}$ have a similar behavior, but with
a larger difference ($\approx$ 30-50\%). The HCO$^{+}$ emission is
strongest for the star with no excess UV (Spectrum C), and is slightly
stronger for Spectrum B than A.

\subsection{NLTE effects on the CN line emission.}
\label{sec: nltechap6}
Special attention should be paid to the CN lines, because they are
among the strongest lines observed in several disks, and because LTE
excitation is not a good approximation for this molecule as shown
below. The integrated intensities of the CN lines are larger by a
factor of 2 to 3 (Table \ref{tab: linesUVdep}) in the models with
excess UV (Spectrum A and B), even though the CN column densities are
a factor of two higher in the outer regions of the disk for the model
without excess UV (Spectrum C) (Fig. \ref{fig: column_Costa4000}).
This appears counterintuitive since the outer disk, due to its larger
surface, should dominate the line emission. This is not the case since
most of the additional CN in the model with Spectrum C is located in
the upper most layers of the disk where the density is so low that the
levels become sub-thermally excited. To show the effect of the NLTE
excitation, the profiles of the CN $3-2$ line for the three stellar
spectra are plotted for NLTE and LTE populations in Fig. \ref{fig:
nltelte}. In the LTE case the emission from the model with Spectrum C
dominates the lines from Spectrum A and B as expected when
considering the column densities (Fig. \ref{fig:
column_Costa4000}). For the NLTE calculations, the order of the
emission reverses since most of the CN column in the model with
Spectrum C is at low density.

\begin{figure}[ht!]
\resizebox{\hsize}{!}{\includegraphics{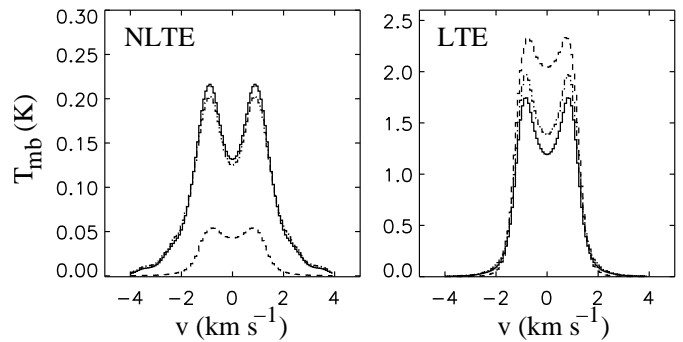}}
\caption{\label{fig: nltelte} Comparison of the CN $3-2$ emission line
for NLTE (left) and LTE (right) populations.  The dash-dotted lines
are for Spectrum A, the solid lines for Spectrum B, and the dashed
lines for Spectrum C.  The spectra have been convolved with a beam of
5.33\arcsec}
\end{figure}
In addition, a few of the hyperfine lines of the $J=1-0$ transition
experience small masing effects due to a population inversion. This
makes NLTE excitation effects important for all CN levels in these
disk environments. The modeling and interpretation of any of the CN
levels should therefore be done with a detailed radiative transfer
study.

\subsection{High resolution simulation of CN/HCN}
\begin{figure}[ht!]
\resizebox{\hsize}{!}{\includegraphics{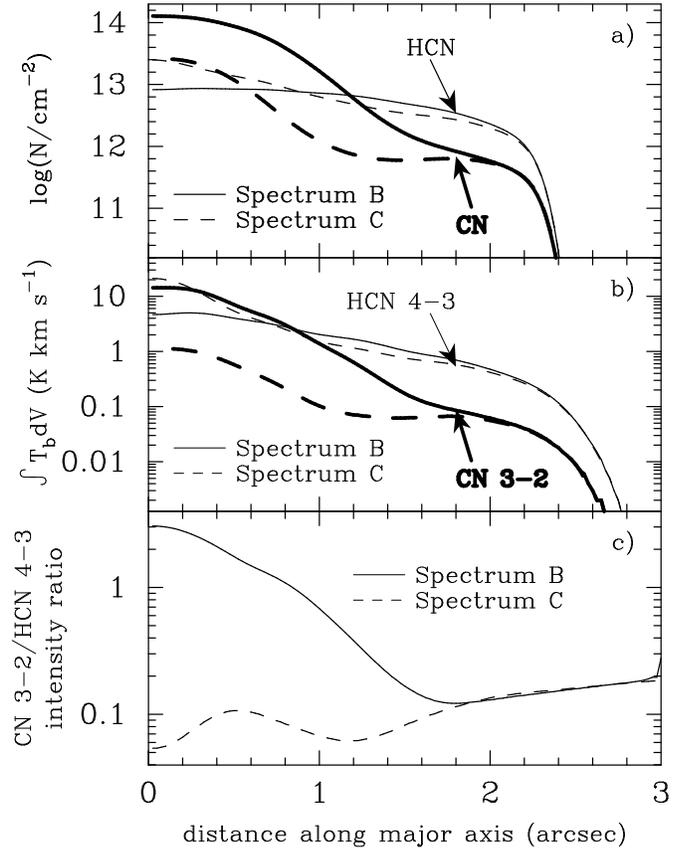}}
\caption{\label{fig: almares} a) The beam averaged column densities of
CN and HCN. b) Integrated intensities for the CN 3--2 and HCN 4--3
transition. c) Ratio of the integrated intensity of CN 3--2/HCN
4--3. The modeled intensities and columns were calculated using a beam
of 0.3\arcsec. The solid lines represents the emission of a star with
stellar UV (Spectrum B), the dashed lines a T-Tauri star with no
excess UV (Spectrum C). The thick lines in the plots (a) and (b)
represent the CN and the thin lines the HCN column-densities and
intensities.}
\end{figure}

The CN and HCN column densities (Fig. \ref{fig: column_Costa4000})
show clear differences for the different radiation fields. Their
abundance ratio in the disk is a sensitive function of radius due to
the different radial gradients of the column densities of CN and
HCN. In the future, telescopes like the Atacama Large Millimeter Array
(ALMA) will be able to probe this ratio spatially. In order to
simulate ALMA observations, the molecular line intensities are
calculated assuming a FWHM gaussian beam of 0.3\arcsec, which is
chosen based on the spatial resolution of the model grid and which
will easily be reached with ALMA.  In Fig. \ref{fig: almares}c the
velocity integrated intensity ratio along the major axis of the disk
is plotted for spectra B and C.
The intensity ratio for spectrum B peaks close to the star and drops
to a 15 times lower ratio at large radii. Spectrum C shows the
opposite behavior rising by a factor 4 from close to the star to
larger radii. The emissions at these large radii are similar
(Fig. \ref{fig: almares}b,c), so that high resolution is essential for
distinguishing the effects of different UV radiation fields. For a
better understanding of the emission profile, the beam averaged
column densities and integrated intensities for CN and HCN are given
in Fig \ref{fig: almares} $a$ and $b$, respectively.  The drop of
intensities at $\sim$ 2.5\arcsec~ is due to the partial filling of the
beam.  In both cases the ratio of the intensities for spectrum B and C
is smaller than the ratio of the column densities by roughly an order
of magnitude; in the inner most region with spectrum C, for example,
the column density ratio of CN to HCN is almost unity , while the
intensity ratio is $\lesssim 0.1$.  The expected sensitivity of ALMA
at 345 GHz in a 0.3\arcsec \ beam is $\sim$0.3 K km s$^{-1}$ in 1
hr. Thus, the HCN and CN distribution in the inner 0.6" can readily be
imaged and the Spectrum B and C models should be easily
distinguishable. Imaging of CN in the outer part of the disk will
require longer integration times.
\section{Discussion}
\label{sec: discussec}

The calculated abundances and column densities show orders of
magnitude variations between different models. The inclusion of ND-2D
UV radiative transfer instead of an approximate 2D transfer decreases
the abundances of species like HCN, CN and CS. When CO dissociation
via stellar UV is taken into account, a large amount of atomic carbon
is produced, which is converted to radicals (CN, C$_{2}$H) and direct
reaction products (CS, HCN). The adopted stellar spectrum has a clear
effect on the column densities of most species except for
CO. Especially the CN and HCN column densities show different slopes
with increasing radius. These variations should be detectable with
sufficient spatial resolution as modeled in Fig. \ref{fig: almares}.

The effect of UV radiative transfer and the stellar UV spectrum
are less clear in the integrated line intensities with larger beams
for the following reasons. First, the telescope beams available, now
and in the near future, average the disk over the entire beam. Second,
most lines observed to date are the brightest lines available. A large
number of these are optically thick; CO for instance is only capable
of probing the upper layers of the disk. Third, some lines suffer from
sub-thermal excitation. The NLTE effects can be severe in the
optically thin upper layers of the disk, while the LTE intermediate
layers are at a large optical depth.
 
A comparison between different millimeter line radiative transfer
codes (van Zadelhoff et al 2002) shows that the standard deviation on
the level populations can reach 12\% for the higher levels in
NLTE. These calculations were performed for an extreme combination of
optical depth ($\tau$) and NLTE. Even though the $\tau$ in disks is
very high, the levels are not as far from LTE as in the above
mentioned calculation. It is therefore safe to assume that the
molecular line calculation in this paper (Fig. 9 and 10) has
`error-bars' similar or less than $12 \%$.  This is smaller than the
error-bars on the observations, which are estimated to be roughly
20\%.

 In Paper I, the model results underestimated the observed molecular
abundances of species such as CN and HCN, and overestimates CO in most
disks. One of the motivations of the present study is to check if the
inclusion of CO dissociation via stellar UV could resolve these
discrepancies.  In comparison to the results presented in Paper I,
both the CN emission and the CN/HCN ratio has significantly increased
using the full-2D UV radiative transfer code. The column density of
C$_2$H is also increased.  However, the CO column density itself is
not much affected by dissociation via the stellar UV. The column
densities of non-radical species such as HCN are almost the same as in
Paper I and are thus lower than observed in some sources.

There are several important physical processes which need to be
investigated in future studies (see also \S 2.1).  The UV transfer was
calculated here using a particular set of grain scattering
properties. The mean intensities derived in the disk change when
different properties are adopted. Since the abundances do not depend
linearly on the mean intensity, they are not expected to differ much
as long as the gas/dust ratio and grains sizes are similar to those in
interstellar clouds. If the dust sizes have increased to mm-sized
grains, this would alter both the dust properties and the dust
settling. This may allow more UV photons to reach deeper into the
disk, enhancing the dissociation of CO.

The inclusion of X-rays could significantly change the chemical
abundances (Aikawa \& Herbst 1999; 2001).  The secondary UV field
generated by the excited H$_{2}$ could result in a higher abundance of
radicals at lower disk heights. This would enhance the total line
emission due to both the higher CN and HCN column densities and the
increased excitation of the molecules at higher
densities. Hydrodynamics such as accretion and turbulent mixing would
smooth out molecular stratification to some extent, and transport of
radical species would affect the chemical processes in neighboring
regions.  On the observational side spatially and spectrally resolved
observations are needed to find unique tracers of different
processes.

\section{Conclusion}

The main conclusions from this work are:

\begin{itemize}
\item{The transfer of stellar ultraviolet radiation has to be treated in
a two-dimensional manner for a correct abundance determination of
radicals like CN and C$_{2}$H.}
\item{The photodissociation of CO by the stellar radiation supplies
additional atomic carbon, which leads to significantly higher
abundances of molecules, especially of radicals like CN and C$_2$H.
CO itself shows no detectable dependence on the stellar radiation
field, making it a poor tracer of the chemical processes in disks.}
\item{The abundances of the molecules in the `radical region' are also
sensitive to the adopted stellar UV spectrum. This effect will not
necessarily be reflected in the current observed line intensities,
however, since the lines may be sub-thermally excited in the surface
layers where the abundances are most enhanced. Spatially resolved
observations by ALMA are needed to distinguish the different
scenarios.}
\item{Observational constraints on the fluxes from T Tauri and Herbig
Ae stars at ultraviolet wavelengths, in particular in the critical
912--1100 \AA \, region, are urgently needed.}
\end{itemize}

{\it Acknowledgments.}  The authors are grateful to the referee
for suggestions and comments that improved the paper considerably.
They also thank P. D'Alessio and V. Costa for sending and discussing
the disk model and stellar UV field used in the paper. GJvZ
acknowledges M.\ Spaans and C.P.\ Dullemond for helpful discussions on
the UV radiative transfer. Astrochemistry in Leiden is supported by a
SPINOZA grant from the Netherlands Organization for Scientific
Research (NWO). Y.A. is supported by the Grant-in-Aid for Scientific
Research on Priority Areas of the Ministry of Education, Science and
Culture of Japan (13011203; 14740130).  The chemical models were
carried out at the Astronomical Data Analysis Center of National
Astronomical Observatory in Japan.

\appendix
\section{Axi-symmetric 2D continuum radiative transfer}
\label{sec: appenA}
A two-dimensional (2D) axi-symmetric Monte Carlo code was set up to
calculate the transfer of ultraviolet radiation by absorption and
scattering due to grains.  In the calculation, the photons are
actually followed in 3D, but by applying axial symmetry the grid can
be rotated at any time in $\Phi$ such that the photon comes back to
the $\Phi$=0 plane. This causes the photons to follow hyperbolic paths
in the 2D $\Phi$=0 plane . The main reason for adopting a 2D rather
than 3D approach is the increase in accuracy in the solution for the
same number of photons or amount of computer time.

\subsection{Photon propagation}
\label{sec: follow}
Each photon is started at the source of radiation with an
intensity given by the boundary conditions for the source. These 
boundary conditions are discussed in Sec. \ref{sec: bound}. A photon
package can undergo two different processes: absorption  and scattering
by dust. A model photon represents the amount of energy in a
wavelength bin, given by the total surface of the source multiplied by
the flux through this surface divided by the number of photons in this
wavelength bin. Per wavelength bin, one needs sufficient resolution to
sample all possible angles of emission. In general the total number of
photons per wavelength bin $\delta \lambda$ is given by

\be
N_{phot,\delta \lambda} = C1 \times n\phi \times n\theta \times nR
\times nz, 
\ee 
\noindent
where $nR$ and $nz$ are the number of cells in the $R-$ and
$z-$directions, respectively, and $n\phi$ and $n\theta$ are the
directional resolutions. One can multiply this by an arbitrary
constant (C1 $\ge 1$) to improve on the statistics. For a reasonably
sized grid ($nR=nz=70$), $10^{6}$ photons suffice in practice per
wavelength bin. In general 34 wavelength bins are used in between 912
\AA \, and 1$\mu m$.

The initial energy stored in each photon package is given by
\begin{equation}
I_{i}(0,\lambda)=\frac{F_{\lambda}\times S}{N_{phot,\delta \lambda }}
\end{equation}
where $F_{\lambda}$ is the flux entering the system and $S$ the total
surface it passes through.

\begin{center}
\begin{figure}[h!]
\resizebox{8cm}{!}{\includegraphics{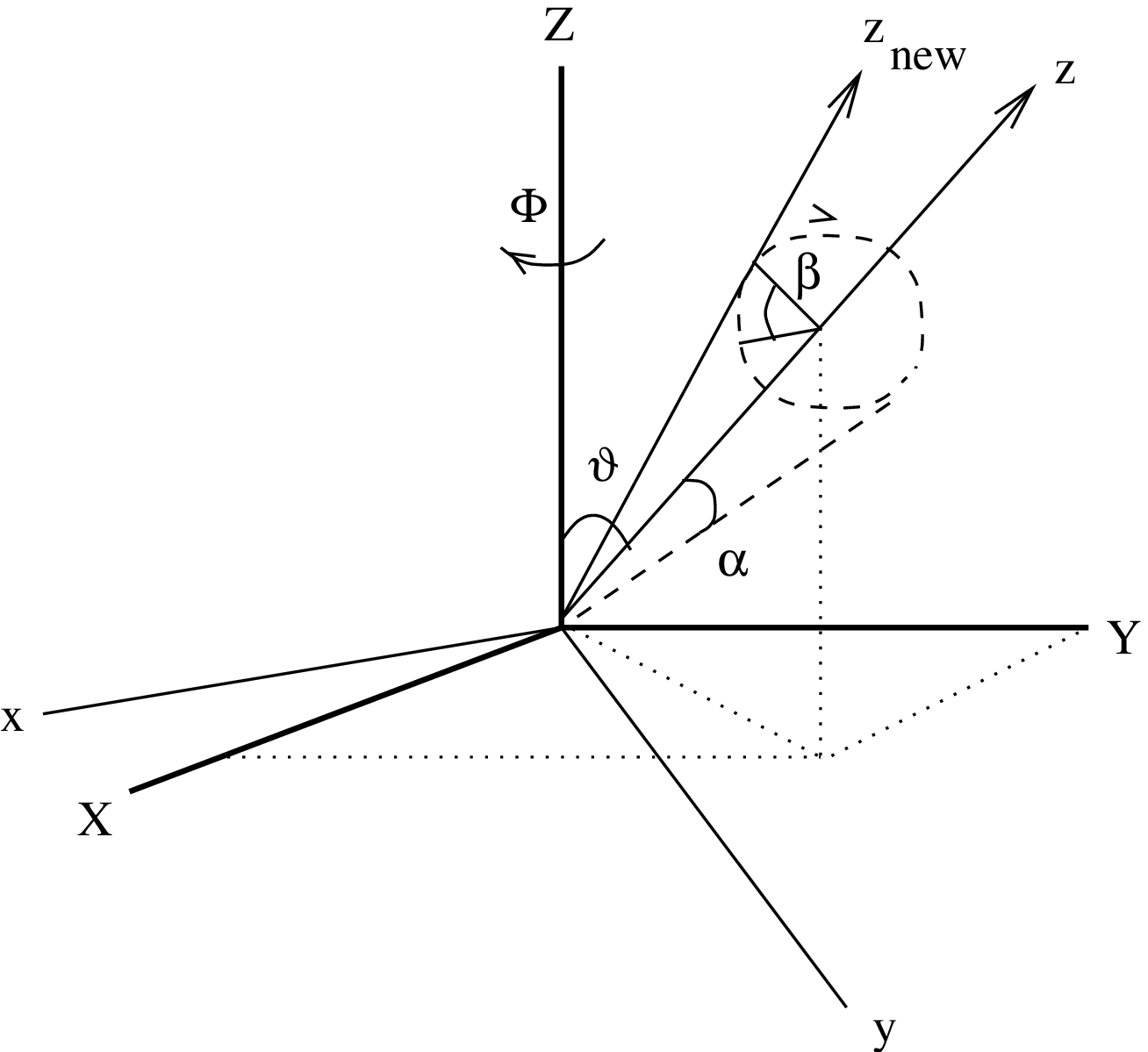}}
    \caption[Direction and scattering of photons]{\footnotesize
    Scattering of a photon propagating in the direction
    ($\phi$,$\theta$) through angles $\alpha$ and $\beta$. The new
    photon direction will lie on a cone defined by $\alpha$ and
    $\beta$.}
\label{fig: scatter}
\end{figure}
\end{center}

At the surface of emission source, the scattering optical depth is
first calculated using a random number generator \be
\tau_{scat}=\rm{-ln}(1-\zeta)
\label{eq: scat1}
\ee
with $\zeta$ a random number between 0 and 1. This optical depth can
be converted to a total absorption optical depth
\be
\tau_{abs}=K_{\lambda} \times \tau_{scat}, 
\label{eq: scat2}
\ee 
where $K_{\lambda}$ is the fraction of absorption compared to
scattering.
The photon package travels through the grid until it
reaches the optical depth ($\tau_{abs}$) at which it scatters. In the
mean time, its intensity drops according to

\ba I_{i}(s+\Delta s,\lambda)&=& I_{i}(s,\lambda) e^{-\Delta
\tau_{\lambda}},
\label{eq: tau}
\ea where $\Delta \tau_{\lambda}$ is related to the physical
properties through \ba \Delta \tau_{\lambda}&=&
\frac{1}{2.5\rm{log_{10}}(e)} \times \Delta \rm{A}_{\lambda} \\ &=&
\frac{gd(R,z)}{2.5\rm{log_{10}}(e)} \times \frac{\Delta s \times
n(\rm{H+2H_{2}})}{1.59 \cdot 10^{21}}\times \frac{\rm{A}_{\lambda}}{\rm{A}_{v}}
\label{eq: const} 
\ea
\noindent
 with A$_{\lambda}$ the visual extinction at wavelength $\lambda$,
gd(R,z) the gas--to--dust ratio compared to 100, $\Delta s$ the path
length traveled within a grid cell and $n(\rm{H+2H_{2}})$ the total
number density in the grid cell. The relation $n$(H)+ 2$n$(H$_{2}$)=
1.59$\Ex{21}$A$_{v}$ is taken from Savage et al. (1977).  When the
photon package reaches the point of scattering, the entire package
changes direction according to the Henyey-Greenstein phase function
(Henyey-Greenstein 1941, Witt 1977): \ba P({\rm
cos}\alpha,g_{\lambda})=\frac{1-g^{2}_{\lambda}}{4 \pi
[1+g_\lambda^{2}-2g_{\lambda}\cos \alpha]^{3/2}}
\label{eq: henyey}
\ea with $g_{\lambda}=<\rm{cos}\alpha>$, the mean scattering
angle. The angles under which the photon scatters are in reference to the
original direction of motion [defined with an ($x,y,z$) coordinate
system, such that $z$ is in the direction of the photon propagation
prior to scattering] \ba \alpha&=&\arccos \left(
\frac{1}{2g_{\lambda}}
\left[(1+g^{2}_{\lambda})-\left(\frac{1-g^{2}_{\lambda}}{1-g_{\lambda}+2g_{\lambda}\zeta}
\right)^{2} \right] \right) \\ \beta&=& \pi (2 \zeta -1)  \ea
\noindent
with $\alpha$ the angle between the $z$-axis and the new direction
(z$_{\rm new}$) and $\beta$ the rotational angle in the $x-y$ plane
(Fig. \ref{fig: scatter}), with $\zeta$ a random number between 0 and
1.
These angles are transformed back to the rest frame $(X,Y,Z)$. After the new
direction is calculated the next step is determined using equations
\ref{eq: scat1} and \ref{eq: scat2}.  The photon continues its path
until it reaches the outer boundary of the computational grid or has a
negligible intensity defined by the user.

\subsection{Computation of the mean intensity}

The mean intensity is determined in each cell through summation of all
photons passing through the grid cell. As each photon represents an
amount of energy which diminishes as it traverses a (dense) medium,
the energy added to the cell is the mean energy per second of the
photon along the path $\Delta s$ traveling with the velocity of
light. The mean energy density $u_{\lambda}(iR,iz)$ of the radiation
field is then calculated by dividing the total summed energy by the
volume $V$ of the gridcell $iR, iz$.  \be
u_{\lambda}(iR,iz)=\frac{1}{V(iR,iz)} \sum I_{i}\frac{\Delta s}{c}
\frac{(1-e^{-\Delta \tau})}{\Delta \tau},
\label{eq: utoj1} 
\ee 
 The mean intensity $J_{\lambda}(iR,iz)$ in the grid cell is directly
related to the mean energy density $u_{\lambda}(iR,iz)$ according to
\be J_{\lambda}(iR,iz)=\frac{c}{4 \pi} u_{\lambda}(iR,iz).
\label{eq: utoj}
\ee 

The mean intensity is the only output needed in this
application, so the above description of scattering
is sufficient. If the code is to be applied for 
both intensity and polarization, the four Stokes parameters
have to be calculated (e.g., Whitney 1991). This is in principle
readily implemented in the code, but has not been done here.

\subsection{H$_{2}$ and CO self-shielding}
\label{sec: shield}
Photo dissociation of H$_{2}$ proceeds through absorptions in the
Lyman and Werner bands in the 912--1100 \AA \, region. About 10-15 \%
of all excitations lead to the dissociation of H$_{2}$. Since this
process proceeds through line excitation the absorption lines become
optically thick once the H$_{2}$ column in each level reaches
$\sim$10$^{14}$ cm$^{-2}$. As this is easily reached for the objects
of interest in this paper, self-shielding of H$_2$ has to be taken
into account. Only a limited amount of continuum radiation can be
absorbed by H$_{2}$, leaving a large flux available for other species
to be dissociated. The equation used to calculate the shielding due to
H$_{2}$ is taken from Draine \& Bertoldi (1996), their equation 37 is
shown here as \ref{eq: bert1}. The equation describes the
self-shielding of the H$_{2}$ molecule through the Lyman and Werner
series: \be f_{s}(\rm{H}_{2})= \frac{0.965}{(1-x/b5)^{2}} +
\frac{0.035}{\sqrt{1-x}} \times e^{-8.5\times 10^{-4} \sqrt{1+x}}
\label{eq: bert1}
\ee 
with x$=\frac{N(\rm{H_{2}}) \, [\rm{cm^{-2}}]}{5\Ex{14}}$ and
$\rm{b}_{5}=\frac{\rm{b} \, [\rm{cm \, s^{-1}}]}{1\Ex{5}}$ and $f_{s}$
the shielding of the UV in the H$_{2}$ lines.  Since H$_{2}$ dissociates
through lines, only a fraction of the continuum in between 912
-- 1110 \AA \, is absorbed.

This equation is valid for each photon package $j$ traveling through a
gridcell since the formulae is defined as the drop in intensity of the
radiation field due to the column of H$_{2}$. First the mean value of
the function $f_{s}$, for each package during its step is calculated,
where the mean is taken over the column density $N$ of the step: \be
\overline{F}(j,iR,iz)= \frac{\int^{N2}_{N1} f_{s}(j,iR,iz)
dN}{\int^{N2}_{N1} dN} \label{eq: mean1} \ee

This then leads to a mean value for the grid cell through: 
\be F_{\rm cell}(iR,iz)=\frac{\sum^{np(iR,iz)}_{j=0} \overline{F}(j,iR,iz) \Delta
s(j,iR,iz)}{\sum^{np(iR,iz)}_{j=0} \Delta s(j,iR,iz)},\label{eq:
mean2} \ee with $np$($iR,iz$) the number of photon steps taken in the
grid cell defined by $iR$ and $iz$ and $\Delta s$ the length of the
step of the photon. The value of $F_{\rm cell}$ is then tabulated for
each cell and applied to the intensities between 912--1100 \AA, which
suffered dust extinction only.

Among the other molecules, only CO is affected significantly by
self-shielding. In the other cases the dust extinction combined with
H$_{2}$ extinction, if the dissociation is at similar wavelengths
(e.g., CN photodissociation or C photoionization), is sufficient to
describe the rates of ionization and dissociation.  CO shielding is
treated in a similar way as H$_{2}$ with the difference that instead
of using an analytical solution, the values are taken from a table by
Lee et al. (1996), in which shielding factor is given as a function of
H$_2$ column density, CO column density and $A_{\rm v}$. Calculation
of the shielding factor thus require the CO abundance in each grid cell.  In
principle, iteration on the chemistry and radiative transfer is needed
to calculate the precise CO dissociation rates. In the case of
circumstellar disks, a large column is reached for each single cell
due to the extremely high total columns, especially in the inner parts
of disks. This causes the stellar dissociation rates to depend only on
the mean value of the CO abundance in each cell.  A constant CO
abundance of CO/H$_{2}$=$7\cdot10^{-5}$ is assumed when calculating the CO
self-shielding.  If one is interested in a medium for which the
precise positional change in the CO shielding is important, the
spatial grid near the boundary at which the flux enters has to be
increased and the chemistry and radiative transfer have to be
iterated. Only a small chemical network is needed for this purpose,
since the CO abundance depends mostly on a few simple molecules and
atoms.

\subsection{Sources of radiation}
\label{sec: bound}
\begin{figure}[ht!]
\begin{center}
\resizebox{\hsize}{!}{\includegraphics{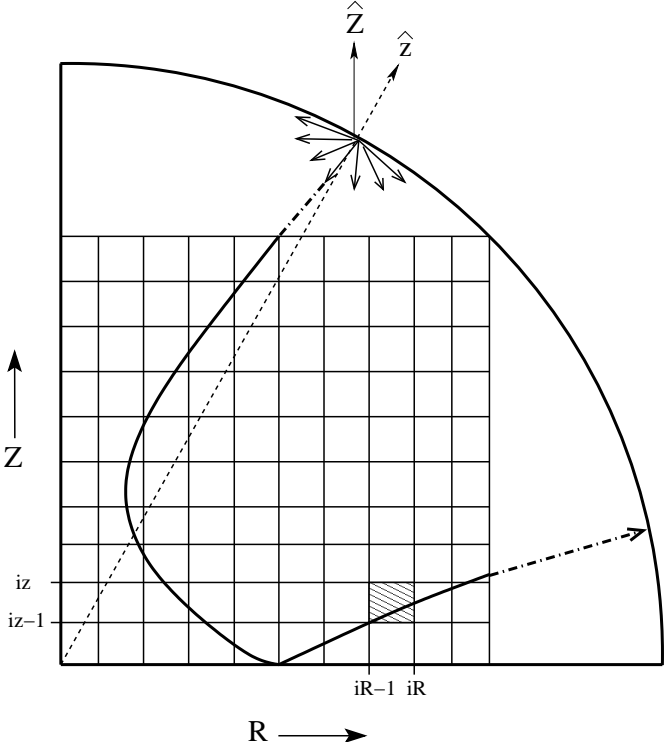}}
\end{center}
\caption[Computational grid]{\footnotesize The grid 
represents the computational domain
with a virtual shell around it. The dashed line denotes the  the frame
to describe the angle distribution of incoming photons from the boundary.
The solid curved line is the
path of the photon-package, the dot-dashed line is its virtual path.}
\label{fig: obslkca}
\end{figure}
Radiation can come from several sources. Discussed here are two
sources of radiation: interstellar radiation and stellar radiation.
Since all boundary conditions consist of a surface through which flux
passes, there is a general description for the angle distribution of
the photon packages. These angles, \be \Theta=\rm{cos}^{-1}
\sqrt{1-\zeta}
\label{eq: begin1}
\ee
\be
\Phi=2 \pi \zeta 
\label{eq: begin2}
\ee are given in the frame in which the $z$-axis lies perpendicular to
the tangent plane of the source surface.  Equations (\ref{eq: begin1})
and (\ref{eq: begin2}) need to be transformed to calculate the angles
in the plane where its $z$-axis is parallel to the $Z$-axis of the
problem (Fig. \ref{fig: obslkca}).  The equations have been taken from
Yang et al. (1995) and describe the isotropic emission from a surface.
\begin{itemize}
\item{{\it Interstellar field:} The interstellar field is assumed to
be isotropic over 4$\pi$ steradian while the computational domain is a
rectangle in $R-z$ space (Fig.~\ref{fig: obslkca}). To calculate the
mean intensity, a virtual sphere with a radius equal to that of the
upper right corner is taken in order to represent the angle
distribution of the incoming field. The starting positions of the
photons are randomly taken at the surface of the sphere with a
direction given by Equations \ref{eq: begin1} and \ref{eq:
begin2}. These directions are in the frame denoted by \^{z}. For the
calculation, the directions are transformed to the computational frame
denoted by \^{Z}.  The photon packages are placed at the edge of the
computational rectangle after which the photons follow the routine
described in Sec. \ref{sec: follow}. At each grid cell $(iR,iz)$
through which the photon passes, the mean intensity is updated. In
general the Draine interstellar radiation field with the van Dishoeck
\& Black (1982) extension for $\lambda > 2000 $ \AA \, is used which
can be scaled by a factor if required.}
\item{{\it Stellar light:} Since the medium in our problems is
generally much larger than the stellar radius, the star can safely be
assumed to be a point source. However, the total energy emitted does
depend on the radius of the star. In this paper we assumed
$R_{\star}$=2$R_{\odot}$ The far-field expression for the mean
intensity in a medium with negligible absorption is, \be
J_{\lambda}(r)=\frac{I_{\lambda}}{4}\left(\frac{R_{\ast}}{r}\right)^{2},
\label{eq: meanint}
\ee where $I_\lambda$ is the intensity at the surface of the star.  It
is a good approximation in the region of $R \gtrsim 5R_{\ast}$, which
we are concerned with.  For problems where the mean intensities close
to the star are important, emission from the real stellar surface
should be taken into account. }
\end{itemize}

\end{document}